\documentclass[aps,pra,preprint,superscriptaddress]{revtex4-1}

\usepackage[latin1]{inputenc}
\usepackage{amsmath}
\usepackage{amsfonts}
\usepackage{amssymb}
\usepackage{graphicx}
\usepackage{xspace}
\usepackage{color}
\usepackage{comment}
\usepackage[hidelinks]{hyperref}

\newcommand{\sSNOM}{\textit{s}-\mbox{SNOM}\xspace}
\newcommand{\figRef}[1]{Fig. \ref{#1}}
\newcommand{\insertFigure}[4]{
\begin{figure}
	\centering
  \includegraphics[width=#3]{#1}
  \caption{\label{#2}#4}
\end{figure}
}

\begin{document}

\author{Benjamin Pollard} 
\thanks{These authors contributed equally to this work.}
\affiliation{Department of Physics, Department of Chemistry, and JILA, University of Colorado, Boulder, CO 80309, USA}
\author{Eric A. Muller} 
\thanks{These authors contributed equally to this work.}
\affiliation{Department of Physics, Department of Chemistry, and JILA, University of Colorado, Boulder, CO 80309, USA}
\author{Karsten Hinrichs}
\affiliation{Leibniz-Institut f\"ur Analytische, Wissenschaften - ISAS - e.V., Department Berlin, Albert-Einstein-Str. 9, 12489 Berlin, Germany}
\author{Markus B. Raschke}
\email{markus.raschke@colorado.edu}
\email{Corresponding author: markus.raschke@colorado.edu}
\affiliation{Department of Physics, Department of Chemistry, and JILA, University of Colorado, Boulder, CO 80309, USA}
\title{Nano-spectroscopic imaging of intermolecular structure, coupling and dynamics}

\begin{abstract}
Molecular self-assembly, the function of biomembranes, and the performance of organic solar cells rely on molecular interactions on the nanoscale.
The understanding and design of such intrinsic or engineered heterogeneous functional soft matter has long been impeded by a lack of spectroscopic tools with sufficient nanometer spatial resolution, attomolar sensitivity, and intermolecular spectroscopic specificity. 
We implement vibrational scattering-scanning near-field optical microscopy (\sSNOM) in a multi-spectral modality with unprecedented spectral precision to investigate the structure-function relationship in nano-phase separated block-copolymers. 
We use a vibrational resonance as a sensitive reporter of the local chemical environment and resolve, with few nanometer spatial resolution and 0.2 cm$^{-1}$ spectral precision, spectral Stark shifts and line broadening correlated with molecular-scale morphologies.
By creating images of solvatochromic vibrational shifts we discriminate local variations in electric fields between
the core and interface regions of the nano-domains with quantitative agreement to dielectric continuum models.
This new nano-chemometric ability to directly resolve nanoscale morphology and associated intermolecular interactions can form a basis for the systematic control of functionality in multicomponent soft matter systems. 
\end{abstract}

\maketitle

Many classes of soft matter gain their unique functionality from intrinsic or engineered inhomogeneities 
on nanometer length scales.
In particular, the phase behavior of multi-component nanoscale molecular composites such as biomembranes, organic photovoltaic systems, liquid crystals, and polymer heterostructures are defined by intra- and intermolecular interactions. 
This intermolecular coupling dictates local morphology and vice versa.
Precisely that complex microscopic interplay, in particular when associated with internal hetero-interfaces \cite{Barbour2007a}, controls the macroscopic properties and performance of a material.

The general challenge and opportunity of controlling rich molecular-scale correlations in functional soft matter systems is exemplified in understanding the amphiphilic self-assembly of block-copolymers (BCPs), which phase separate into a wide range of domain textures on macro-molecular (3 to 150 nm) length scales \cite{Knoll2002}.
Many potential applications for photonic crystals, lithographic templates, and engineered membranes are limited by nanoscale defects and disorder in the self-assembled morphologies of BCPs \cite{Gopinadhan2013}.
The phase separation of BCPs results in different local densities, degrees of crystallinity, and chemical environments for specific functional groups between the center and interface of a polymer block domain.

Several established techniques such as grazing incidence X-ray diffraction (GIXRD), small/wide angle X-ray scattering (SAXS/WAXS) \cite{Wang2011}, near-edge X-ray absorption fine structure (NEXAFS) microscopy \cite{Ade2009}, transmission electron microscopy (TEM), electron tomography, and various scanning probe microscopies \cite{Giridharagopal2010} give access to nanoscale morphology in soft matter.
Despite these advances, simultaneous insight into the underlying spatial inhomogeneity of intermolecular coupling, charge transfer, and their corresponding dynamics has remained difficult.

In that regard, vibrational spectroscopies, such as vibrational Raman \cite{Carach2012a} and linear or multidimensional infrared \cite{Barbour2007},
 are exquisitely sensitive to local chemical 
 environments. 
However, these techniques in their conventional implementation do not provide nanoscale spatial information and
only indirectly characterize the correlation of local morphology with molecular coupling \cite{Pensack2009}.
Vibrational spectroscopic characterization is thus needed with chemical selectivity, high spectral precision, attomolar sensitivity, and nanometer spatial resolution on the morphological length scales which determine the function of soft-matter systems \cite{Shima1994,Guest2011,Peter2008,Wubbenhorst2003}.

Here, we perform nanometer-scale chemical imaging of domain morphology with simultaneous spectroscopic insight into underlying intermolecular coupling and associated dynamics. 
We achieve nanoscale chemical contrast through infrared vibrational scattering scanning near-field optical microscopy (\sSNOM).
Furthermore, using the high spectral power and resolution of a tunable quantum cascade laser, we extend IR \sSNOM to true multispectral imaging (two spatial, one spectral dimension) in order to probe spectral variations as a function of spatial BCP morphology with unprecedented 0.2 cm$^{-1}$ spectral precision at $\sim$10 nm spatial resolution.

We perform
spatio-spectral mapping, using carbonyl resonances to locally probe the differences in intermolecular interactions at interfaces and within the nano-domains of quasi-lamellar PMMA in thin films of PS-{\em b}-PMMA.  
From spatial maps of the spectral position and lineshape of the carbonyl mode, we gain insight into local chemical environments  of sub-ensembles within the phase separated morphology.
Furthermore, spectroscopic imaging within the amorphous PMMA domains enables the discrimination of effects due to electrochemical heterogeneity native to a non-crystalline polymer from effects resulting from miscibility (such as Stark shifts) near PS:PMMA domain boundaries.

\figRef{Fig1}(a) shows the layout of our experimental setup.
Mid-infrared light from a quantum cascade laser (QCL, Daylight Solutions) tunable between 1660 - 1900 cm$^{-1}$ with 0.2 cm$^{-1}$ precision is focused onto the apex of a Pt coated tip (ARROW-NCPt, NanoWorld AG) of a modified atomic force microscope (AFM, Innova, Bruker) using an off-axis parabolic mirror (NA=0.4, effective focal length 19.5 mm), with a power density of $\le$$50$ MW/cm$^2$. 
Tip scattered near-field signal is detected in a confocal epi-illumination/detection geometry with a HgCdTe detector (MCT, KLD-0.25/DC/11.5, Kolmar Technologies), and far-field background is suppressed by a lock-in filter (HF2, Zurich Instruments) demodulated at the third harmonic of the tip tapping frequency $\omega_c$.
Tip scattered light is recombined at the detector with light of known phase from the reference arm $\phi(\bar{\nu}_{ref})$, in an asymmetric Michelson interferometer geometry. 
The near-field amplitude $A(\bar{\nu})$ and phase $\phi(\bar{\nu})$ are simultaneously determined by modulation of the reference phase at frequency $\Omega$, 
similar to \cite{Ocelic2006}. 
$A(\bar{\nu})$ and $\phi(\bar{\nu})$ can then be related to the complex refractive index of the sample, as will be discussed below.
For comparison, each sample was also measured using far-field ellipsometry, yielding ensemble-averaged spectra of the real $n(\bar\nu)$ and imaginary $k(\bar\nu)$ components of the complex index of refraction \cite{Rosu2009}.
\insertFigure{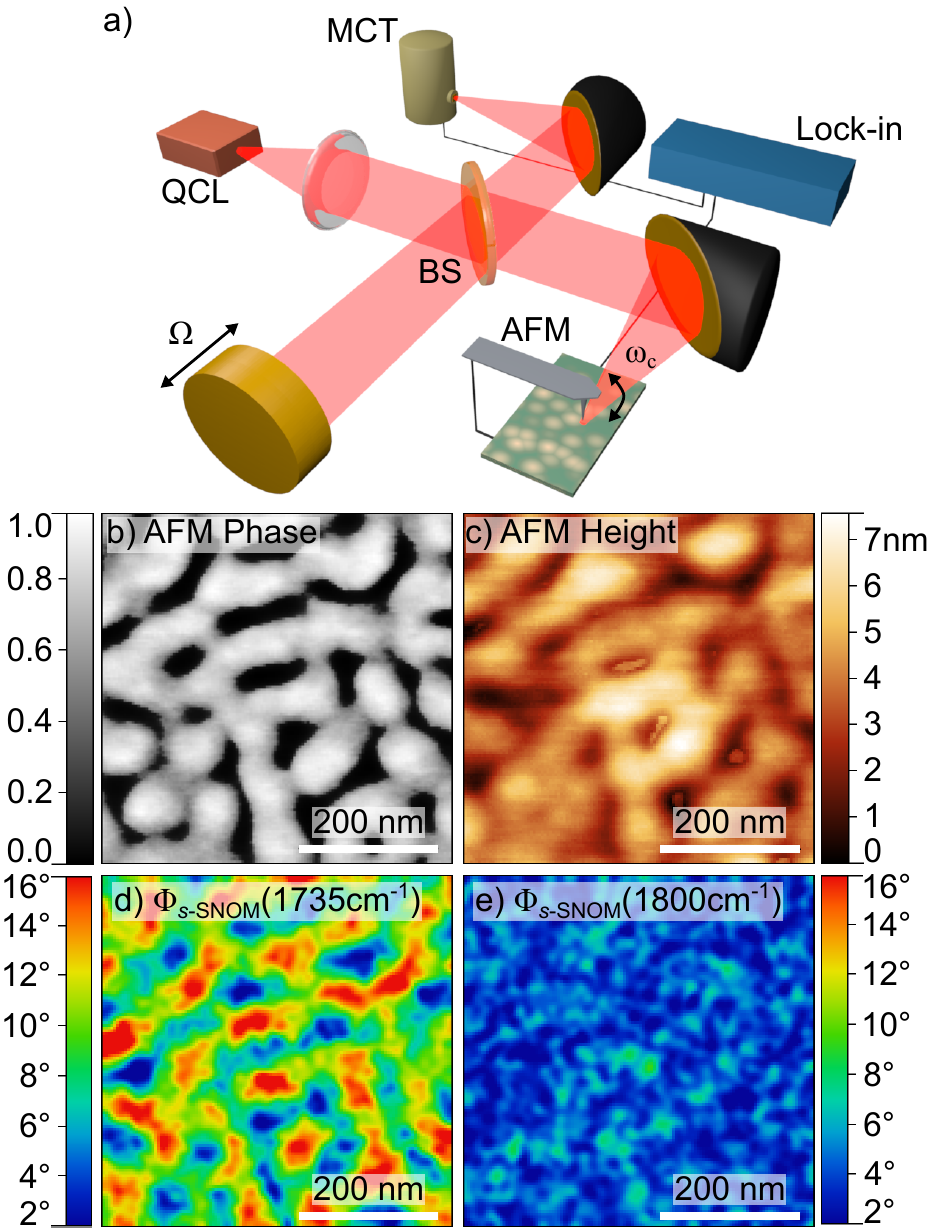}{Fig1}{9.5cm}{
(a) Diagram of interferometric \sSNOM experiment for spatio-spectral nano-imaging, with atomic force microscope (AFM), tunable quantum cascade laser (QCL), MCT IR detector, beamsplitter (BS), and lock-in amplifier. 
Tip oscillation at $\omega_c$~$\sim$250 kHz, and reference arm mirror oscillation at $\Omega$ $\sim$500 Hz.
Sample scans of quasi-lamellar PS-\emph{b}-PMMA showing AFM tapping phase (b, arb. units), AFM height (c), and near-field phase ($\phi_{\sSNOM}$) on (d) and off resonance (e) with the carbonyl mode (4-pixel Gaussian filter). 
}

A high molecular weight block copolymer of polystyrene-block-poly(methyl methacrylate), (PS-{\em b}-PMMA), with a relative chain length of 270.0-{\em b}-289.0 $M_n \times 10^3$ (P4443-SMMA, Polymer Source) is spin coated from a 1\% w/v solution in toluene onto native-oxide silicon substrates (1-4 kRPM). 
Long chain lengths result in large irregular quasi-lamellar structures with incomplete phase separation \cite{Walheim1997}. 

As shown in the AFM images, \figRef{Fig1}(b-c), differences in the viscoelastic properties between PMMA and PS result in contrast between nanodomains of the two polymers, in tapping phase and topography \cite{Knoll2001}.
\figRef{Fig1}(d-e) show
simultaneously recorded near-field
images of the \emph{s}-SNOM phase with photon energy resonant (d) and non-resonant (e) with the carbonyl vibrational mode of PMMA. 
The strength of the \emph{s}-SNOM phase signal is proportional to the resonant response of the sample, and can thus be used to map the local density of carbonyl groups.

\insertFigure{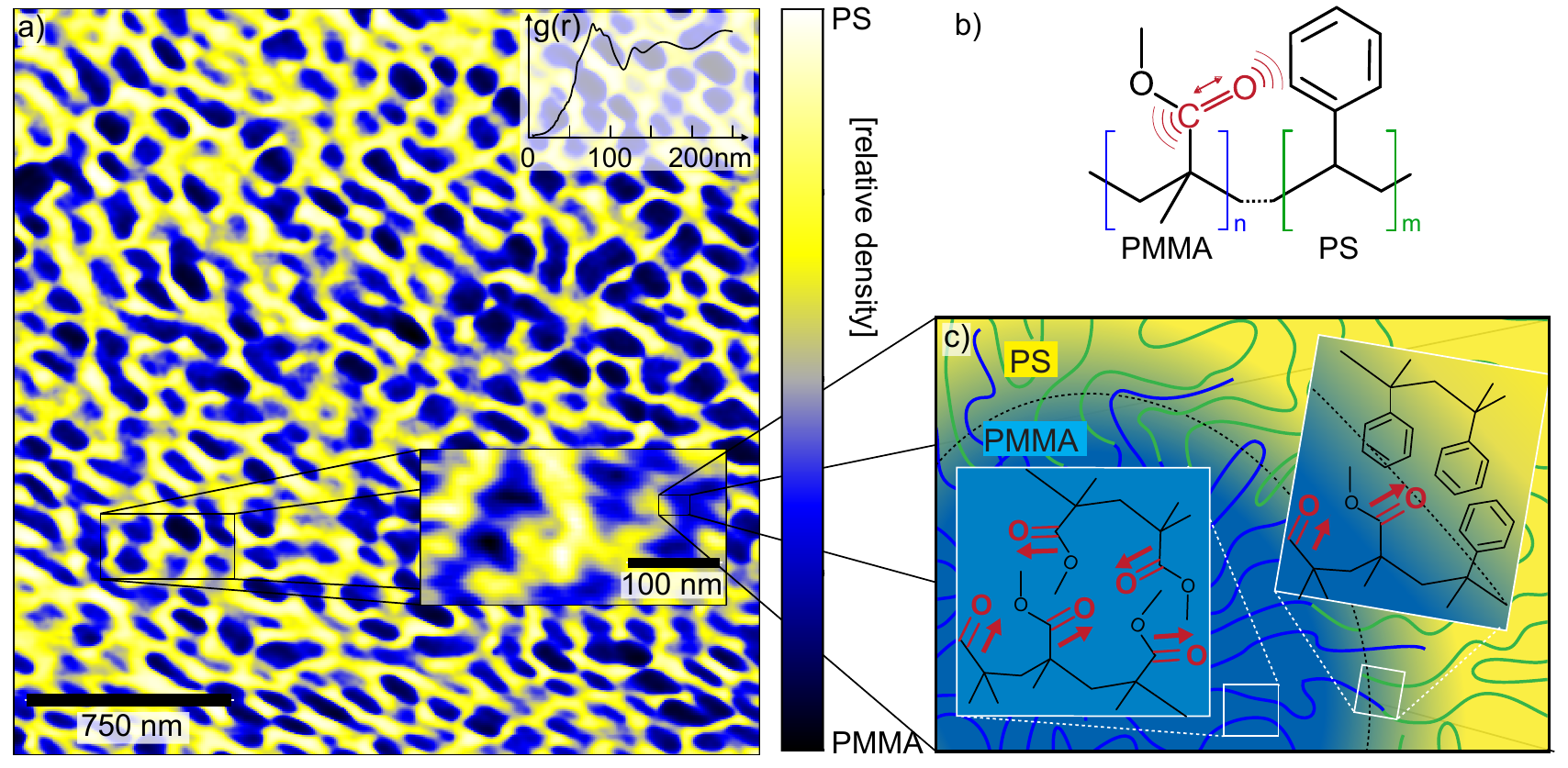}{Fig2}{17.35cm}
{
(a) Nanoscale chemical map of self-phase separated amphiphilic block-copolymer PS-{\em b}-PMMA derived from carbonyl resonant \sSNOM imaging.
(inset) The radial distribution function $g(r)$ is obtained from the large area scan.
(inset) High spatial resolution (330nm width) \emph{s}-SNOM image showing additional features and continuous variation in relative PMMA:PS density across the interface.
(b) Molecular structure of (PS-{\em b}-PMMA), highlighting the carbonyl marker resonance.
(c) Schematic of domain morphology of PS-{\em b}-PMMA with incomplete segregation near the interface. 
}

Because the carbonyl resonance is found in PMMA and not PS (\figRef{Fig2}(b), the \sSNOM phase is linearly related to the relative density of PMMA:PS within the $\sim10^3$~nm$^3$-scale near-field probe volume of the tip \cite{Raschke2003a}.
\figRef{Fig2}(a) illustrates the chemical contrast observed in a large area \sSNOM scan of PS-b-PMMA. 
Spatial Fourier analysis reveals a radial distribution function $g(r)$ with a peak at $80 \pm 20$ nm (\figRef{Fig2}(a), top inset), indicative of the length scale of microphase separation. 
A high spatial resolution image (\figRef{Fig2}(a), bottom inset) shows additional morphological details within the $\sim$10~nm resolution limit.
The relative density of PMMA in the high resolution image is observed to vary across the nanodomain interface, over a typical length scale of 30-40~nm, as a result of the incomplete phase separation.
As schematically shown in \figRef{Fig2}(c), the associated spatial variation in local chemical environment are expected to give rise to observable changes in the spectroscopic signatures of the carbonyl oscillator as a reporter mode,
between the center and interfacial region of the domains.

We then acquired multispectral \sSNOM optical phase images of a 500~nm~$\times$~225~nm region of a quasi-lamellar PS-\emph{b}-PMMA film by sequential sample scanning (\figRef{Fig3}(a)).
We imaged the carbonyl resonance from $1660-1800$~cm$^{-1}$, incrementing by 3~cm$^{-1}$ for each scan. 
The characteristic carbonyl resonance gives rise to a continuous variation of  \sSNOM spectral phase $\phi(\bar{\nu})$ and amplitude $A(\bar{\nu})$ when self-referenced to non-resonant PS (which exhibits a low and flat response in this spectral range).  
\figRef{Fig3}(b) shows the resulting \emph{s}-SNOM spectrum of the nanoscale PMMA region indicated in (a).
From the molecular density of PMMA of $6.2$ monomers/nm$^3$ and an near-field signal depth of $\le 10$ nm corresponding to first order to the lateral spatial resolution \cite{Raschke2005a,Taubner2005}, this spectrum corresponds to a sub-ensemble of only 25 zeptomole or $\sim 10^4$ carbonyl oscillators.
The peak center is slightly shifted compared to far-field ellipsometry due to the spectral phase approximation \cite{Xu2012} with only negligible effects from tip-sample coupling
\footnote{The large off-resonant contributions to the dielectric function in the infrared spectral range give rise to a comparably weak on-resonant dispersion. 
This results in $\phi_{s\rm -SNOM} \propto k(\bar{\nu})$ to a good approximation \cite{Xu2012}.
While the imaginary part of the {\em s}-SNOM signal more directly resembles $k(\bar{\nu})$ \cite{Huth2012}, in polar phase and amplitude, rather than cartesian Re and Im signal representation, phase uncertainty does not affect the lineshape (see \cite{Xu2013} for details). 
Furthermore, direct measurement of the phase minimizes susceptibility to external fluctuations (such as laser noise).
}.

\insertFigure{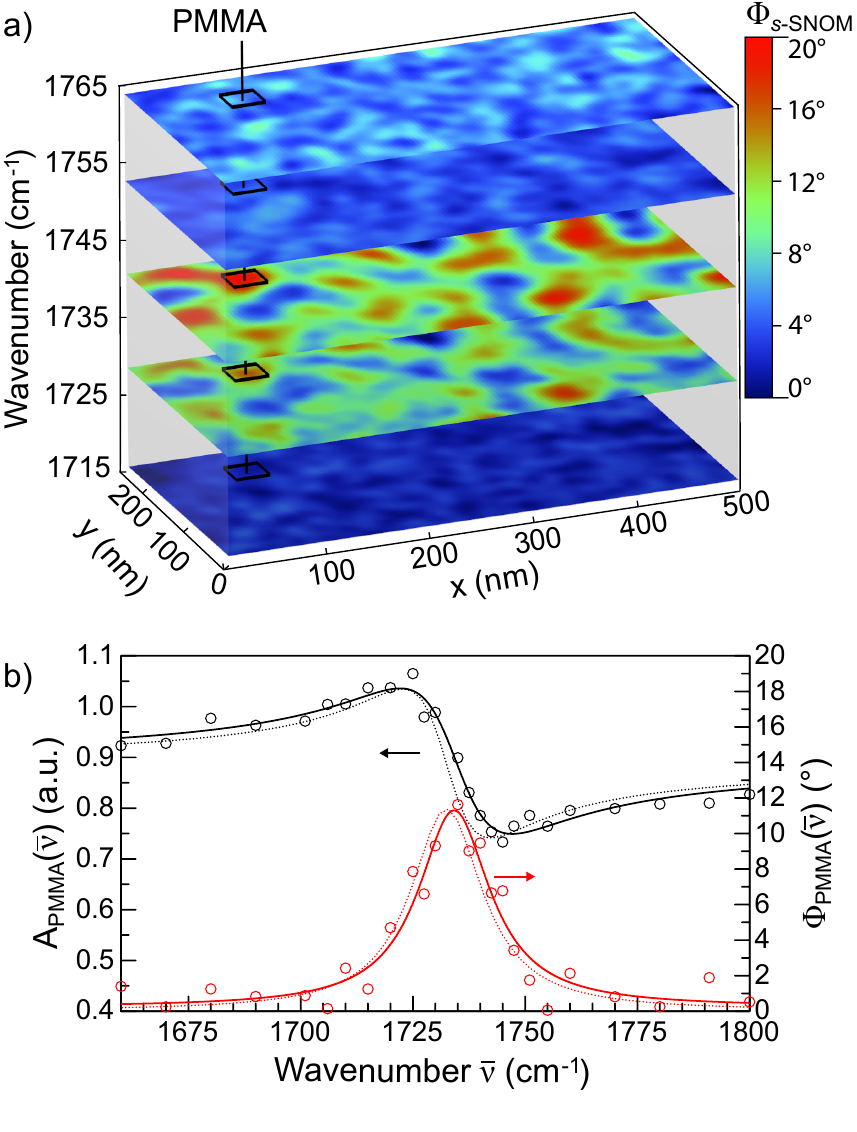}{Fig3}{8.3cm}
{(a) Multispectral \sSNOM phase image of PS-{\em b}-PMMA. Carbonyl resonant PMMA region highlighted for illustration.
(b) \sSNOM amplitude $A(\bar{\nu})$ and phase $\Phi(\bar{\nu})$ spectrum of a $\sim$25$\times$25 nm PMMA region (referenced to a non-resonant PS region of similar size), with real and imaginary Lorentzian fits for the amplitude and phase, respectively. Dotted lines show $n(\bar{\nu})$ (black) and $k(\bar{\nu})$ (red) from spectral ellipsometry, scaled vertically to match the near-field data.
}

The analysis of the nanoscale spectral phase $\phi(\bar{\nu})$ spectra in terms of peak characteristics gives access to nanoscale variations in local chemical environment.
\figRef{Fig4} shows the result of Lorentzian line fits from the complete multispectral scan, with extracted peak position $\bar{\nu_0}$ (a) and FWHM linewidth $\Gamma$ (b) for each 10 nm $\times$ 10 nm pixel area.
In general a spectral red shift and broader linewidth is found towards the center of PMMA domains, compared to a blue-shifted and narrower spectral response at the domain interface
\footnote{While there is a lower oscillator density and thus overall lower optical phase response at domain interfaces, Lorentzian fitting effectively decouples the amplitude and peak location from the spectral lineshape. 
Some artificial shifts through the spectral phase approximation from varying signal strength may occur, but they are negligibly small and if present would contribute a shift in the opposite direction compared to what is measured.}.
Despite some spatial inhomogeneities in the amount of resonance shifts and line broadening, a 
clear difference between spectra within domain centers compared to interfaces is evident when considering the ensemble distribution of pixels over the entire spatio-spectral image.
As shown in \figRef{Fig4}(c), a correlation plot of $\bar{\nu}_0$ vs. $\Gamma$ separates points of varying carbonyl oscillator density with associated spatial regions by the integrated spectral phase (colorbar).
It shows the trend of redshifting and broadening
towards the domain centers, which are characterized by higher carbonyl oscillator density. 
As a specific example, \figRef{Fig4}(d) shows the distinct center versus interfacial behavior across a single domain covering a 60 nm distance as indicated in \figRef{Fig4}(a) and (b)(black line). 
The peak is observed to monotonically redshift and broaden towards the center of the domain, and blueshift and narrow towards the interface.

\insertFigure{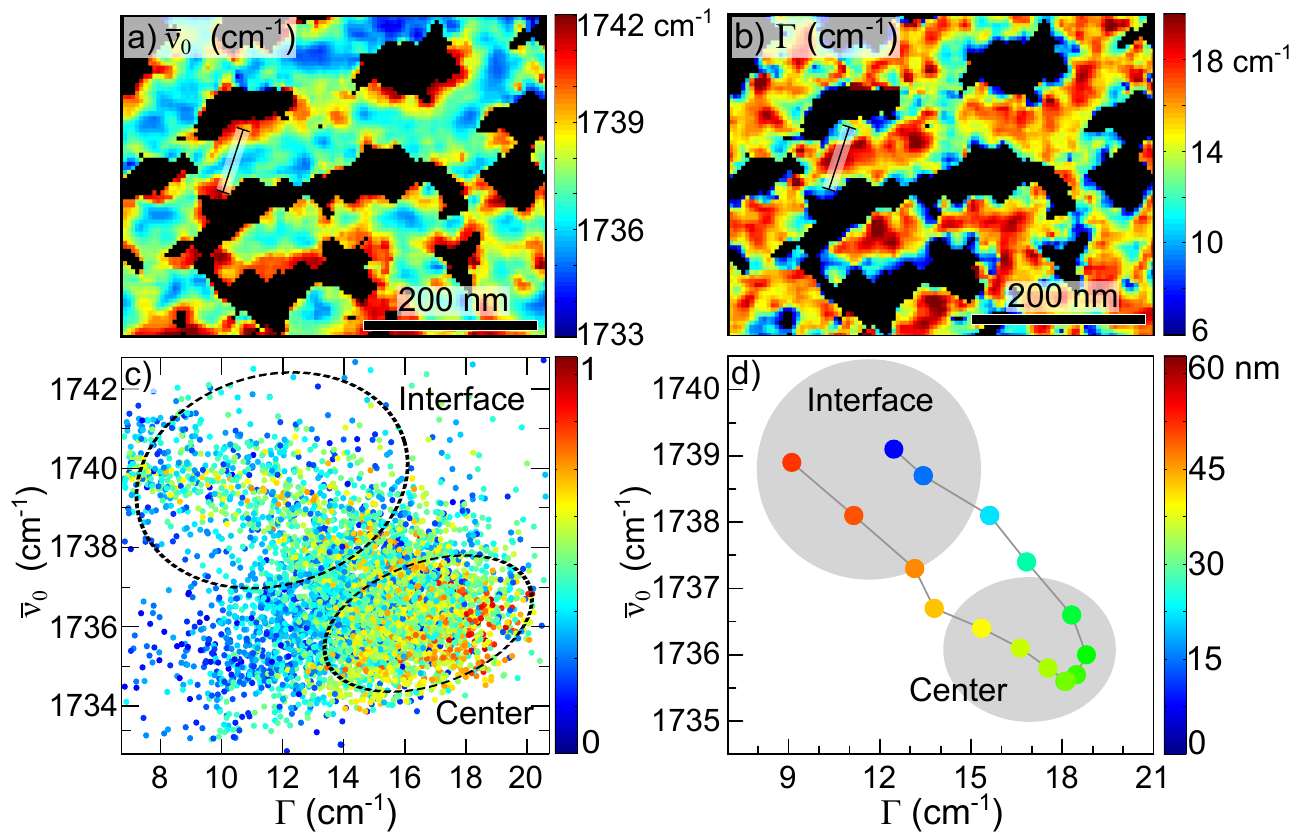}{Fig4}{12.5cm}
{(a-b) Spatio-spectral analysis of carbonyl peak center $\bar{\nu}_0$ (a) and linewidth $\Gamma$ (b) of measured \sSNOM phase for each 10 nm $\times$ 10 nm scan region. 
Black areas could not be fit to resonant lineshape, due to a too low PMMA density in the PS-dominated regions.
(c) Scatterplot of $\bar{\nu}_0$ vs. $\Gamma$ variation of the data from (a) and (b), with color code referring to spectrally intensity $I$, integrated over the area of the peak in phase, corresponding to the oscillator density of PMMA. 
(d) Plot of $\bar{\nu}_0$ vs $\Gamma$ for individual points along the 60~nm spatial profile indicated by the black line in (a) and (b), exemplifying the overall trend shown in �, and illustrating the morphological trend underlying the observed spectral variation.
}


We attribute the red-shift towards the center of PMMA domains to a spatially-varying Stark shift, caused by heterogeneity in the local chemical environment. 
Stark effect induced red-shift ${\Delta}{\bar{\nu}_{0}}$ of the carbonyl mode is proportional to the local electric field $\langle{\vec{F}}\rangle$ given by:
\begin{equation}
\label{StarkCoeff}
{h{c}{\Delta}{\bar{\nu}_{0}}}=-{\Delta}{\vec{\mu}_{probe}}{\cdot}{\vec{F}}.
\end{equation}
A change in dipole moment upon photoexciation,  ${\Delta}\vec{\mu}_{probe}$, also termed the Stark tuning rate, results in a red shift of the reporter mode with increasing field strength \cite{Levinson2012}. 
Carbonyl modes have a large Stark tuning rate, and thus enable the methyl-ester group of the PMMA polymer to act as a 
particularly sensitive reporter of local electric field strength. 
Thus, the observed red-shift of the carbonyl vibration towards the center of the PMMA  domains indicates larger local electric fields within the domains relative to their interface.

These local electric fields result from the electrostatic energy of solvation within and between polymer chains.
Solvation and solvatochromism of polar molecules can be described by dielectric continuum models, which have successfully been applied 
to quantitatively describe, for example, local structure and intramolecular electric fields in proteins \cite{Levinson2012}.  
In the Onsager model of solvation, which  describes electric fields within the solvation shell surrounding a polar solute molecule, the local electric field ${\vec{F}}$ surrounding a dipolar solute is proportional to the dipole moment of the molecule $\vec{\mu}$, modified by its refractive index $n$, and the dielectric constant $\epsilon$ of the solvent \cite{Onsager1936},
and given by
\begin{equation}
\label{RField}
{\vec{F}}={\frac{\vec{\mu}}{a^3}} \Bigg [\frac{2({\epsilon}-1)(n^2+2)}{3(2{\epsilon}+n^2)}\Bigg],
\end{equation}
where $a$ is the solvation shell radius of the solute molecule and can be approximated from bulk densities.
Within this model, we can describe the solute probe as a segment of 
a syndiotactic PMMA chain with a dipole moment 
oriented along the carbonyl 
bond of the methacrylate group (\figRef{Fig1}(b-c) red arrows). 
Solvatochromism of its methyl-ester stretch results from changes in the local dielectric environment, due to mutual self-interaction of neighboring methyl-ester groups of PMMA.
Because of the large chain length in our
large molecular weight block copolymer, incomplete phase separation and significant mixing occurs near the interface, and the local dielectric constant is a continuously varying function of PS:PMMA ratio across the interface 
region.
The dielectric constant of PS is lower than that of PMMA, and thus red-shifts are expected with increasing local density of PMMA relative to PS.
\insertFigure{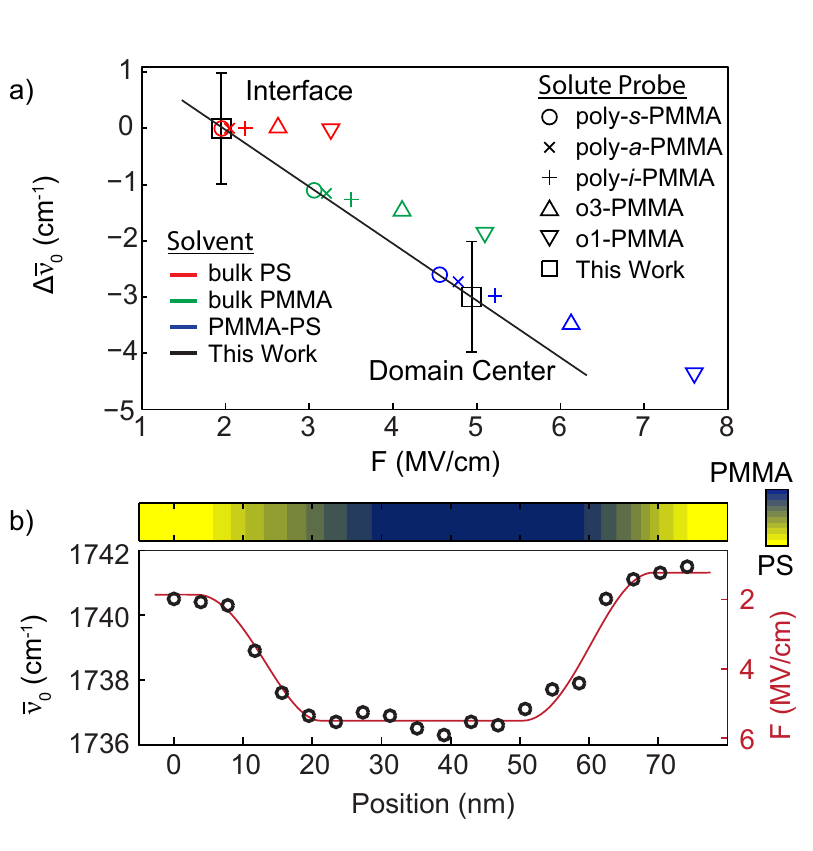}{Fig5}{8.3cm}
{(a) Modeled PMMA solvatochromic spectral shifts in various dielectric environments referenced against the shift predicted for a molecular probe in bulk PS.  
Electric field and Stark shift are calculated for dielectric environments in PS (red), bulk PMMA (green) \cite{Zhou2004}, and PMMA nano-domains in a PS-PMMA blend (blue) \cite{Hamanoue1983}. 
The molecular probe can be approximated by bulk dielectric parameters for various tacticities of PMMA, syndiotactic- (circle) , atactic- (X), or isotactic- (+) PMMA chains, as well as for shorter oligomer segments (trimer o3-PMMA) and the methacrylate monomer (o1-PMMA). 
Local \emph{s}-SNOM measurements of the Stark shift (cm$^{-1}$) at the domain centers relative to the interface for syndiotactic-PMMA:PS are indicated by black squares. 
The spectral shifts are aligned with Onsager fields calculated using bulk dielectric parameters assuming a solute probe consisting of polymeric \emph{s}-PMMA.
(b) Spatially resolved spectral shift observed in a line scan across a single domain, and its good correlation with the local relative concentration of PMMA (inset colormap as measured by the resonant \emph{s}-SNOM phase).
\label{Onsager}
}

As seen from the correlation plot in \ref{Fig4}(c), the carbonyl mode redshifts by $\sim$3~cm$^{-1}$ measured in the PMMA dense domain centers, relative to the PS rich inferface regions. 
Using literature values for the dielectric properties of PMMA and PS, we compare our results to predictions from the Onsager model.
Figure \ref{Fig5}(a) shows that the local electric field surrounding the carbonyl mode is expected to increase from $\sim$2 MV/cm in PS to $\sim$5 MV/cm in PMMA.
Based on these modeled local electric field strengths between center and interface, and the range of observed shifts in \ref{Fig4}(c),  we derive, using equation \ref{RField}, a Stark tuning rate of $\sim$1~cm$^{-1}$/(MV/cm). 
This value is consistent with literature values \cite{Pensack2010,Gearba2011}.

The microscopic variations of the experimentally measured Stark shift span a broad distribution, yet a comparable distribution of local electric field strengths can be predicted through the Onsager model, as shown in Figure~\ref{Fig5}(a).
This distribution occurs because bulk properties of a polymer depend heavily on structural details of the polymer chain rather than simply the monomeric chemical identity.
Different stereocomplexes of PMMA, isotactic (\textit{i}-PMMA), atactic (\textit{a}-PMMA), and syndiotactic (\textit{s}-PMMA) which are used here differ only in chirality, yet \textit{i}-PMMA has a larger mean-square dipole moment ${\langle}{\mu^2}{\rangle}$ due to a larger degree of chain order.
Measured bulk values of ${\langle}{\mu^2}{\rangle}$ are also heavily dependent upon chain length; monomers (o1-PMMA) or oligomers (trimeric o3-PMMA) of PMMA have larger ${\langle}{\mu^2}{\rangle}$ and thus are expected to undergo larger Stark shift in the same local solvent environment as compared to longer chains \cite{Ando1997}.
Morphologically induced ordering also affects dielectric properties, and PMMA nano-domains in PS-PMMA copolymers have been observed to have unusually large dielectric permittivity as evidenced by large Stark shifts of a dye molecule probe \cite{Hamanoue1983}.
Rotation of individual monomers, local fluctuations in density \cite{Shima1994}, and increased chain mobility within 10-50 nm of a surface also affect the dielectric properties\cite{Peter2008,Wubbenhorst2003}. 
Variation in local chemical environment is thus expected for nanoscale sub-ensembles of a polymer blend with molecular scale disorder.
The continuum description thus acts as a lower bound for predicted spectral shifts.
The spectral shifts observed in \emph{s}-SNOM thus quantitatively agree with the model of solvation induced DC Stark shift. 
Indeed the seemingly unaccounted for
spatial inhomogeneities in measured spectral shifts and linewidths 
can be expected from known variation and distributions of bulk dielectric properties in PMMA.

Futher microscopic insight can be obtained from mapping the peak center of the carbonyl stretch  as a line scan across a single nanoscale domain (\figRef{Onsager}(b), simultaneously with the PMMA local density as obtained from the resonant \emph{s}-SNOM phase (inset).
Using the value for the Stark tuning rate of $\Delta\mu$=1~cm$^{-1}$/(MV/cm), we can translate spatial changes in peak center to the reaction field (red). 
While the peak shift in general correlates with mixing ratio, the Stark shift depends also on other structural details as discussed above.

Heterogeneity induced Stark shifts have been recently observed in a number of polymer and thin film systems with far-field spectroscopies \cite{O'Brien2011}.
In particular, time-resolved infrared spectroscopy in poly-3-hexyl-thiophene / phenyl-C61-butyric acid methyl ester (P3HT/PCBM) bulk heterojunctions \cite{Pensack2010}, found a dynamic red-shift in the methyl ester stretch of PCBM following charge separation, which was attributed to higher field strengths in PCBM domain centers compared to interfaces.
Conversely, annealing induced changes in morphology of a P3HT/PCBM resulted in Stark shifts of the methyl-ester stretch as observed via  linear FTIR, which was attributed to increased electric field strengths at the interface (rather than domain center) as a result of increased conductivity and decreased screening lengths\cite{Gearba2011}.
With our spatio-spectral implementation of infrared \emph{s}-SNOM, we overcome difficulties in assigning mophologically induced spectral features. 
We are able to image and thus directly correlate morphology and spectral signatures, conclusively assigning red-shifts and broadening to a local chemical environment originating at domain centers. 

In addition to spectral red-shifts, we find a correlation between vibrational linewidth and position relative to domain boundaries, with line broadening in the domain center, as seen in \figRef{Fig4}(c-d).
We believe that this is a result of both increased and inhomogeneous solute-bath coupling, and the availability of additional nearly-degenerate and low-energy modes that contribute to faster vibrational dephasing (when compared to the domain interface with its generally low density and thus weakly coupled carbonyl resonances).
Strength of solvation is commonly correlated with both red-shift of the solute mode and spectral broadening as a result of increased system-bath coupling \cite{DeCamp2005}. 
Competing and opposing contributions to linewidth, however, can complicate such a simple picture. 
Contributions to homogeneous and inhomogeneous broadening result from a complex interplay of attractive and repulsive intermolecular forces \cite{Chandler1981}, coupling of the vibrational mode and molecular dipole moment to the bath as a function of density and viscosity \cite{Lynden-Bell1977}, and temperature effects in activated vibrational energy transfer\cite{Shelby1979}.
Inverse correlation between solvent polarity and spectral linewidth in linear FTIR measurements has been observed in cases where increases in vibrational linewidths due to vibrational dephasing are overwhelmed by narrowing of slower dynamical process involved in spectral diffusion \cite{Brookes2013}. 
In PS-{\em b}-PMMA our results suggest that increased intra- and interchain coupling within the PMMA domains seems to increase both vibrational linewidth and Stark shift relative to the PS-rich interfacial regions.

In summary, we have implemented multispectral IR-vibrational nanoimaging as a nano-chemometric tool to spatially resolve intermolecular interaction with nanometer spatial resolution, attomolar sensitivity, and sub-wavenumber spectral precision.
The approach is based on vibrational linewidth, lifetime, and resonant frequency being exquisitely sensitive to intra- and intermolecular interactions, opening direct spectroscopic access to nanoscale bulk and nano-interfaces in heterogeneous multicomponent soft matter. 
It thus provides direct molecular level insight into the elementary processes linking morphology with intra- and intermolecular coupling and resulting macroscopic materials properties.
Infrared-vibrational nano-chemometrics with full spatio-spectral analysis and its possible extension to ultrafast, time-domain implementations can thus provide deep insight into degrees of disorder, crystallinity, strain, stoichiometry, charge transfer, solvatochromism, and other processes which underlie the structure-function relationship of essentially all soft-matter systems.

\section*{Acknowledgements}

We thank Rob Olmon, Ian Craig, and Greg Andreev for valuable discussions and support at various stages of the experiments. 
This work is dedicated to Chris Armacost from Daylight Solutions, who generously provided the QCL laser, and who tragically passed away during the course of this work. 
Funding is gratefully acknowledged from the National Science Foundation (NSF CAREER Grant CHE 0748226). 
Samples were prepared under a partner proposal at the Environmental Molecular Sciences Laboratory (EMSL), a national scientific user facility from DOE's Office of Biological and Environmental Research at Pacific Northwest National Laboratory (PNNL). PNNL is operated by Battelle for the US DOE under the contract DEAC06-76RL01830.

\bibliographystyle{naturemag_modBP}
\bibliography{references}

\end{document}